# Quantum measurement driven by spontaneous symmetry breaking

Masahiro Morikawa *and* Akika Nakamichi(*)

*Department of Physics, Ochanomizu University,*
*2-1-1 Otsuka, Bunkyo, Tokyo,112-8610, Japan*
*and*
*(*)Gunma Astronomical Observatory*
*6860-86, Takayama, Agatsuma, Gunma 377-0702, Japan*

The measurement process in quantum mechanics is usually described by the von Neumann projection postulate, which forms a basic constituent of the laws of quantum mechanics. Since this postulate requires the outside observer of the system, it is hard to apply quantum mechanics to the whole Universe. Therefore we propose that the quantum measurement process is actually a physical process associated with the ubiquitous mechanism of spontaneous symmetry breaking. Based on this proposal, we construct a quantum measurement model in which the von Neumann projection is described as the dynamical pro-coherence process. Furthermore, the classically distinguishable pointer parameter emerges as the c-number order parameter in the formalism of closed time-path quantum filed theory. We also discuss the precision of the measurement and the possible deduction of the Born probability postulate.

## 1  Introduction

The standard formalism of quantum mechanics is composed from the two kinds of time evolutions which are qualitatively different with each other. One is the unitary dynamical process described by the Schrödinger equation. The other is the non-unitary measurement process which is described by the von Neumann projection postulate. The former process is deterministic and is uniquely described only by specifying the initial condition, while the latter process is essentially probabilistic and directly introduces the statistical nature into quantum mechanics.

This kind of dual-structure[1] of quantum mechanics causes many technical disadvantages in its applications. This is most prominent in cosmology in which any objective device for the measurement cannot be found anywhere anytime[i]. Actually this fact makes the quantum origin of density fluctuations[2] ambiguous and makes the wave function of the Universe[3] entirely meaningless. Therefore it would be natural to explore the unification of the above two kinds of time evolutions, especially in the form that the latter projection process is described within the generalized form of the Schrödinger time evolution. This necessity for the unification is not restricted within the field of cosmology but actually many physicists in various fields have explored the similar issue[4].

---

[i] An exceptions are ourselves which appeared at the last moment of the long history of the Universe. A theoretical formulation strictly based on the principle of quantum mechanics would be possible, in which the big wave function of the Universe evolves according to the Schrödinger equation and is finally measured by us and materialized. However it destroys the whole description of the standard scenario of cosmology, which is firmly based on the realism or the classical picture.

Most of the attempts so far to clarify the quantum measurement process only deal with (I) the quantum entanglement of the system and the measuring apparatus, and (II) the quantum decoherence process of the full system in which the off-diagonal interference term disappears. However the von Neumann postulate claims more. Actually there are further two kinds of processes which have not been commonly discussed. One is the (III) dynamical appearance of c-number degrees of freedom, which distinguish the various quantum states of the system just after the measurement. Another one is the (IV) final transition to a pure state just after the measurement reflecting the fact that the von Neumann postulate is a projection. The states realized in the process IV must have firm correlation with the c-number values in the process III for the faithful measurement. These four processes (I)-(VI) form the *basic quartet* for the measurement process. This is summarized in Table 1.

| PROCESSES | DESCRIPTION |
|---|---|
| I. Quantum Entanglement (=QE) | The states become entangled keeping the initial condition of the system. $\|\psi\rangle \otimes \|\phi\rangle \to c_1\|\psi_1\rangle\|\phi_1\rangle + c_2\|\psi_2\rangle\|\phi_2\rangle$ |
| II. Quantum Decoherence (=QD) | QD reduces the mixing term between the above two states. $\to \rho = \|c_1\|^2\|\psi_1\rangle\|\phi_1\rangle\langle\psi_1\|\langle\phi_1\| + \|c_2\|^2\|\psi_2\rangle\|\phi_2\rangle\langle\psi_2\|\langle\phi_2\|$ |
| III. Classical Order parameter (=CO) | Finite c-number order parameter appears. $\to$ value $\varphi_+$ or value $\varphi_-$, but not both. |
| IV. Quantum Pro-coherence (=QP) | The system becomes a pure state $\to \|\psi_1\rangle$ or $\|\psi_2\rangle$ with the firm correlation with the c-numbers in III. |

**Table 1. The *basic quartet* for the quantum measurement process. The standard construction of the measurement process is composed from the system S and the measurement apparatus A and implicit environment E, i.e. S+A+(E). In the table, each process is described for the simplest two-level state system. The total system initially prepared as $\|\psi\rangle \otimes \|\phi\rangle$, where $\|\psi\rangle = c_1\|\psi_1\rangle + c_2\|\psi_2\rangle$ is the system state, and $\|\phi\rangle$ is the state of the measurement apparatus. Only the process I is described by a unitary evolution equation while all the others II, III, and IV are non-unitary evolutions. In the literature only the processes I and II have been relatively well studied. On the other hand, the processes III and IV form the essential part of the von Neumann projection postulate.**

More precisely about the process III, the common description invoke the c-number property of the pointer system to the introduction of the super-selection rule or the many Hilbert spaces[5]. Although this is reasonable since the completion of the quantum measurement requires the firm storage of information on the quantum state of the system, this static description is not enough. Because we attempt to describe the measuring dynamical process, we need a *dynamical realization of the c-number features*.

It is apparent from the beginning that an apparatus with finite degrees of freedom is impossible to implement the measurement process according to the von Neumann's uniqueness theorem of Hilbert space. We need infinite degrees of freedom that requires the use of quantum field theory. However within the ordinary quantum field theory, the most general description using the effective action only yields the c-number equation of motion for the transition amplitude, which is the operator $\hat{O}$ averaged by the in-vacuum and the out-vacuum $\langle 0_{in}|\hat{O}|0_{out}\rangle$. Since the boundary conditions are set both on the initial in-state and on the final out-state, the evolution equation is not causal. The equation even includes intractable imaginary part in the case of general environment. We have to reformulate the quantum field theory to incorporate an appropriate boundary condition. Such formulation is generally possible[6] and has been applied in various fields of physics[7]. Moreover for our purpose, further development of the formalism would be necessary to describe the evolution equation of the c-number order parameter with dissipation and fluctuation[8]. The present paper is based on this formalism.

More precisely about the process IV, the common argument on the quantum measurement ignores this process despite the fact that the von Neumann projection, after the decoherence process, obviously makes the system a pure state. This is partially correct because the ordinary description of the probability in quantum mechanics is for an ensemble of the copies of the system, and in this case, the prediction based on the density matrix just after the decoherence and that based on the pure state with appropriate probability weight are the same with each other. However, these two methods, density matrix and pure state, deduce different predictions for the multiple measurement or continuous measurement[1]. Thus this process IV is considered to be a basic ingredient of the quantum measurement process.

It is also apparent that this non-unitary evolution in the processes III and IV cannot be described only by a deterministic evolution equation such as simple Schrödinger equation. There must be any mechanism which selects, with appropriate probabilities, one among many possible equivalents. A very similar physical evolution to this would be the process of the spontaneous symmetry breakdown (SSB) during phase transitions. The mechanism SSB is ubiquitous in various fields of physics such as solid state physics, super fluid, elementary particle physics, cosmology, and so on. Therefore it would be natural to consider that such universal process SSB must construct an important ingredient of quantum mechanics as well as the Schrödinger equation. Thus we introduce the hypothesis; the measurement process III and IV always accompanies SSB. In Table 2, the measurement process and the SSB process are compared with each other in various aspects.

| NATURE | QUANTUM MEASUREMENT (III&IV) | SSB PHASE TRANSITION |
|---|---|---|
| Local | Individual measurement ($\Psi$) | A single domain (single phase) |
| Global | Measurement on ensemble ($\rho$) | Many domains |
| Describe local process | von Neumann projection postulate | Langevin equation |
| Describe Global process | Effective equation for density matrix | Fokker-Plank equation |
| Classical variables | Reading of a pointer variable | order parameter (new degrees of freedom emerged) |
| Irreversibility | Yes. It's a projection. | Yes. It is a 1st order phase transition. |
| Meta- stability | Entangled states | Meta-stable states |
| Switching nature | von Neumann postulate | Random force triggers the phase transition |
| Correlation length | Maximum size for the cluster property to hold | Determines the domain size |
| Formation of structures | Generation of classical fluctuations | cosmological inflation, nucleation, spinordal decomposition, topological objects |
| Inter-phase dynamics | fluctuations of pointer values. | Goldstone mode promotes the symmetry recovery. |

**Table 2.** Comparison of the measurement process III&IV and the SSB phase transition process. Both have many common features.

As is shown in detail in Table 2, both the quantum measurement process and SSB process have many common features. Therefore it will be natural to develop the physical measurement process by applying the present general framework of SSB. It should be kept in mind simultaneously that the formalism of SSB at present is not complete and this fact sets the limitation to describe the quantum measurement process in our approach. Our goal in this paper is not the complete resolution of the SSB dynamics but to elucidate the essence of the quantum measurement in the ubiquitous SSB process.

In our approach, the triple role of the c-number random filed, arising from the environment, is conspicuous. Firstly it makes the system decohere in the measurement process II. This kills the quantum fluctuations and induces statistical fluctuations. Secondly it triggers the SSB phase transition in the measurement process III. It is interesting, despite the existence of random fields, that the pure state quantum nature is finally recovered in the system through the phase transition process with positive feedback mechanism, as we will explain the detail in subsequent sections. Thirdly the appropriate strength of the c-number random field is essential to optimize the efficiency of the quantum measurement on an ensemble as we will see in the latter part of this paper.

The construction of this paper is as follows. In section 2, we derive the classical dynamics of SSB in the quantum field theory, especially the emergence and evolution of the c-number degrees of freedom in the closed time-path formalism. This consists of the measurement process III. In section 3, we analyze the quantum dynamics of a system in the environment, especially the evolution from mixed to pure state is discussed as well as quantum decoherence. This consists of the measurement process IV. In section 4, based on the arguments of sections 2 and 3, we introduce a simple model of a single quantum measurement. In section 5, based on this model, the precision of the measurement on an ensemble of the identical copies of the system is studied. In the last section 6, we conclude our study and examine the possibility of the application to cosmology and that of the deduction of the Born probability postulate in quantum mechanics.

## 2  Dynamics of SSB in quantum field theory – emergence and evolution of c-number degrees of freedom (process III)

Let us consider the dynamics of SSB phase transition for clarifying the measurement process III. Here the phase transition dynamics means the existence of the c-number degrees of freedom whose development from zero to non-zero value describes the symmetry breaking dynamical process. In this context, the most appropriate tool would be the effective action $\Gamma[\varphi]$ which is a quantum analog of the classical action functional of the c-number field $\varphi$, where $\varphi$ collectively represents general filed. The effective action $\Gamma[\varphi]$ is defined, in the ordinary quantum field theory, to be the Legendre transformation of the generating functional

$$Z[J] = \langle \Psi_f, t_f | T \exp\left[i \int d^4 x J(x) \hat{\phi}(x)\right] | \Psi_i, t_i \rangle \tag{1}$$

as

$$\Gamma[\varphi] = -i \ln Z[J] - \int d^4 x J(x) \hat{\phi}(x) \tag{2}$$

and

$$\begin{aligned}\varphi(y) &= -i(\delta/\delta J(y)) \ln Z \\ &= \langle \Psi_f, t_f | T \phi(y) \exp\left[i \int d^4 x J(x) \phi(x)\right] | \Psi_i, t_i \rangle,\end{aligned} \tag{3}$$

where $|\Psi_i, t_i\rangle, |\Psi_f, t_f\rangle$ are respectively the initial and final states. These definitions lead to the inverse relation,
$$-J(y) = \delta\Gamma(\varphi)/\delta\varphi(y), \qquad (4)$$
which can be regarded as the evolution equation for $\varphi$ [9]. However, the boundary condition in Eq. (1) makes the equation of motion Eq.(4) non-causal and the quantity Eq.(3) more like a transition amplitude but not the order parameter. Moreover, the quantity Eq.(3) becomes a complex field in general even if we consider a real quantum field.

If the initial and final states $|\Psi_i,t_i\rangle, |\Psi_f,t_f\rangle$ belong to the same Fock space in the ordinary quantum field theory, we can make the generating functional state-independent
$$Z[J] = \langle 0 | T \exp\left[i \int d^4 x J(x) \hat{\phi}(x)\right] | 0 \rangle. \qquad (5)$$
after the $-i\varepsilon$ prescription. However we cannot expect this case in our present issue because what we need now is the dynamical evolution from symmetric to non-symmetric states and $|\Psi_i,t_i\rangle, |\Psi_f,t_f\rangle$ belong to the different Fock spaces.

On the other hand, the static limit of the effective action $\Gamma[\varphi]$, multiplied by a minus sign, yields the effective potential $V_{eff}[\varphi]$ and its minima represent different kinds of vacua characterized by different values of $\varphi$. Each vacuum constructs an independent Fock space which is in-equivalent with each other and cannot be connected by any unitary transformations. Effective action $\Gamma[\varphi]$ is the dynamical analog of $V_{eff}[\varphi]$, and its minima is expected to give some information on the non-unitary dynamics, inter-connecting the above inequivalent Fock spaces.

A minimal extension of the effective action formalism, free from the above mentioned difficulties of causality and complexity, would be the closed-time-path (CTP) formalism[6][7][8], in which the boundary condition is altered. Not only resolving the difficulties but this method properly describes effective dissipation and diffusion as well as quantum corrections in the ordinary quantum field theory as we will see shortly.

The CTP formalism has an extra time-branch and therefore all the time integral is doubled:
$$\int_{-\infty}^{\infty} dt \to \int_C dt, \qquad (6)$$
where the time contour C is depicted in Figure 1; from $-\infty$ to $+\infty$, and then back to $-\infty$.

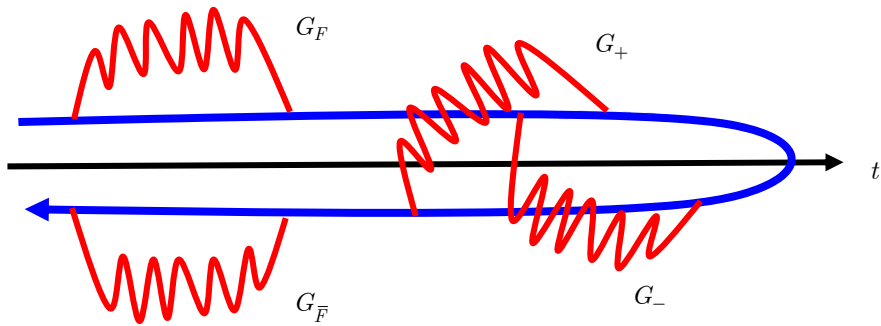

**Figure 1. The time integration contour in the CTP formalism is illustrated in blue curve. The contour C is from $-\infty$ to $+\infty$, and then back to $-\infty$. The components of the two-point green function Eq.(8) are also illustrated in wiggly curves.**

The generating functional in CTP formalism is defined as

$$\tilde{Z}[\tilde{J}] \equiv \text{Tr}[\tilde{T}(\text{Exp}[i \int_C d^4x \tilde{J}(x)\tilde{\phi}(x)]\rho)] \equiv \text{Exp}[i\tilde{W}[\tilde{J}]] \tag{7}$$

where $\tilde{T}$ is the time ordering operator along the extended time-contour C. All the over-tilde means that the quantities are defined on this extended time contour C. The density matrix $\rho$ is the initial state at $t = -\infty$. For example the two point function is also doubled:

$$\tilde{G}(x,y) = -i \begin{pmatrix} \text{Tr}[T\phi(x)\phi(y)\rho] & \text{Tr}[\phi(y)\phi(x)\rho] \\ \text{Tr}[\phi(x)\phi(y)\rho] & \text{Tr}[\bar{T}\phi(x)\phi(y)\rho] \end{pmatrix} = \begin{pmatrix} G_F(x,y) & G_+(x,y) \\ G_-(x,y) & G_{\bar{F}}(x,y) \end{pmatrix}. \tag{8}$$

where $T, \bar{T}$ are the time and anti-time ordering operators, and $F, \bar{F}$ are the ordinary Feynman, anti-Feynman two point functions, respectively. The four components originate from the possible combinations of two branches of the contour C and two arguments $x, y$. Among the four components, only three are independent with each other. This becomes most apparent in the momentum representation:

$$G_F(k) = \frac{D(k) - iB(k)}{D(k)^2 + A(k)^2}, \quad G_{\bar{F}}(k) = \frac{-D(k) - iB(k)}{D(k)^2 + A(k)^2}, \quad G_\pm(k) = -i\frac{D(k) \mp iB(k)}{D(k)^2 + A(k)^2}. \tag{9}$$

The independent functions $D(k), A(k)$, and $B(k)$ have the following meaning[8]. The function $D(k)$ describes quantum corrections such as the renormalization of wave function, mass etc., the function $A(k)$ describes friction which violates the time reversal symmetry, and the function $B(k)$ describes diffusion which induce quantum decoherence effects.

The ordinary formulation of quantum field theory Eqs.(1)-(4) is also possible in CTP formalism. Especially the c-number field defined by

$$\tilde{\varphi}(x) = \begin{pmatrix} \varphi_+(x) \\ \varphi_-(x) \end{pmatrix} \equiv \frac{\delta \tilde{W}}{\delta \tilde{J}} = \text{Tr}[\tilde{\phi}(x) \exp\left[i\int_C d^4x \tilde{J}(x)\tilde{\phi}(x)\right]\rho] \tag{10}$$

has suitable properties as an order parameter. The suffix $\pm$ denote that the arguments are located in the forward time contour ($x_0 \in [-\infty, \infty]$) or the backward time contour($x_0 \in [\infty, -\infty]$), respectively. The effective action is again defined to be the Legendre transformation of $\tilde{W}[\tilde{J}]$.

In this construction, the effective action $\tilde{\Gamma}[\varphi]$ becomes complex in general even if the classical action is real. This situation is the same as the ordinary quantum field theory. However in the present CTP formalism, the imaginary part of the effective action can be uniquely transformed into the kernel of statistical fluctuations. Actually[8], we can rewrite $\tilde{\Gamma}$ as

$$e^{i\tilde{\Gamma}[\varphi]} = \int [d\xi] P[\xi] \text{Exp}[i\left(\text{Re}\tilde{\Gamma} + \int d^4x \xi(x)\varphi_\Delta(x)\right)], \tag{11}$$

where we have introduce an auxiliary field $\xi(x)$. We have defined $\varphi_\Delta = \varphi_+ - \varphi_-$ and $\varphi_C = (\varphi_+ + \varphi_-)/2$. The exponential of the imaginary part of $\tilde{\Gamma}$

$$e^{-\text{Im}\tilde{\Gamma}} = \exp\left[-\frac{1}{2}\iint d^4x d^4y \varphi_\Delta(x) B(x-y) \varphi_\Delta(y) + O(\varphi_\Delta^3)\right] \tag{12}$$

is functionary Fourier transformed into the kernel of the statistical fluctuations for the random field $\xi(x)$

$$P[\xi] = \text{Exp}[-\frac{1}{2}\iint d^4x d^4y \xi(x) B^{-1}(x-y) \xi(y) + O(\xi^3)], \tag{13}$$

which is the Gassian statistical weight upto the second order term in $\varphi_\Delta$ in Eq.(12). In the above, $B(x)$ is the Fourier transform of $B(k)$, and $B(x)^{-1}$ is that of $1/B(k)$. Thus we can now identify the field $\xi(x)$ as a c-number statistical variable with the weight $P[\xi]$ given by Eq.(13). Actually if we define the statistical weight as

$$\langle \cdots \rangle_{\text{st}} \equiv \int [d\xi] \cdots P[\xi], \tag{14}$$

then the correlation of $\xi$ is expressed by the kernel $B$ as

$$\langle \xi(x)\xi(y) \rangle_{\text{st}} = B(x-y). \tag{15}$$

Furthermore, this field $\xi(x)$ couples to the c-number order parameter $\varphi_c$ through the real action on the exponent in Eq.(11). Since the field $\varphi_c$ is already a c-number, the system is considered to possess some objective value at any time of further measurements.

We give here some brief comments on the appearance of the c-number order parameter $\varphi_c$ in the CPT quantum field theory (see [8] for further detail). We emphasize the validity to identify the field $\xi(x)$ as a c-number statistical variable in the following three points. Firstly, our formalism also holds in the system of many harmonic oscillators[10], and is thought to be a quantum filed theoretical generalization of Ref.[10]. The diffusion term in Ref.[10], which represents statistical fluctuations, arizes from the real part of the environment kernel $\text{Re}[\alpha]$, which exactlly corresponds to $\text{Im}\tilde{\Gamma}$ in our formalism. Therefore the function $\text{Im}\tilde{\Gamma}$ should harbor the whole information of the statistical fluctuations. Secondly, we point out the existence of the inverse for the functional fourier transformation, from $\exp[-\text{Im}\tilde{\Gamma}[\varphi_\Delta]]$ to $P[\xi]$ as in Eqs.(12)(13), i.e. the map $\varphi_\Delta \mapsto \xi$ is one-to-one onto. Therefore the functional $P[\xi]$ inherits the whole statistical nature from $\text{Im}\tilde{\Gamma}$. Thirdly, the use of the field $\xi$ is one possible representation of the statistical fluctuations represented by $\text{Im}\tilde{\Gamma}$, and there may be other possible one-to-one onto representation using other field $\xi'$. However they are statistically equivalent with each other since the statistical property is fundamentally governed by $\text{Im}\tilde{\Gamma}$.

As an example in $\lambda\phi^4$ model in the environment with temperature $T$, the imaginary part of the effective action of order $\lambda^2 T$ arises reflecting the information loss to the environment and the uncontrollable energy input from the environment[11]. Even in the vacuum, the imaginary part arises[8]

$$\text{Im}\tilde{\Gamma}[\tilde{\varphi}] = \int \frac{\lambda^2}{128\pi}\theta(|m^2| - \frac{\lambda}{2}\varphi_C^2)\varphi_C^2(x)\varphi_\Delta^2(x), \tag{16}$$

reflecting the instability of the system for $\varphi_C^2 \leq \frac{\lambda}{2}|m^2|$ and according to the above formalism, the c-number field $\varphi(x)$ arises which couples to the c-number random field $\xi(x)$ with the correlation

$$\langle\xi(x)\xi(y)\rangle_{\text{st}} = \frac{\lambda^2}{128\pi}\varphi_C^2(x)\delta^4(x-y). \tag{17}$$

These random fields in general trigger the SSB phase transition.

On the other hand the remaining real part $\text{Re}\,\Gamma + \xi\varphi_\Delta$ of the effective action is considered to be the genuin dynamical action including $D(k)$ and $A(k)$ corrections. Application of the least action principle for the variable $\varphi_\Delta(x)$ yields the Langevin equation for the order parameter $\varphi_C(x)$:

$$\partial_x\partial^x\varphi_C(x) + V'(\varphi_C(x)) + \int_{-\infty}^{t}dt'\int dx' A(x-x')\varphi_C(x') = \xi(x) \tag{18}$$

where the second term $V(\varphi)$ in the LHS includes the renormalized effects from the kernel $D(k)$, the third term includes the dissipative effects from the kernel $A(k)$, and the RHS reflects the diffusive effects from the kernel $B(k)$ through Eq.(15).

In order to elucidate the essence of the process in a technically tractable form, we introduce some approximations to make the system simplest and non-trivial. First we can apply the slow-rolling non-relativistic approximation since the slowly evolving first-stage of the phase transition is relevant to determine the fate of the order parameter. Further we consider the spatially uniform limit of the order parameter since we are interested in the single domain which corresponds, according to the Table 2, to a single measurement of the system. Then the above Eq.(18) reduces[ii] to

---

[ii] The kernels $A, B$ include the information of $\rho$ and so do the parameter coefficients $\gamma, g, \varepsilon$. Such reduction is possible only when the environment is stationary. Moreover these parameters

$$\frac{d\varphi(t)}{dt} = -\frac{1}{\kappa}\left(m^2\varphi(t) + \frac{\lambda}{3!}\varphi(t)^3\right) + \frac{1}{\kappa}\xi(t) \equiv \gamma\varphi(t) - g\varphi(t)^3 + \eta(t). \tag{19}$$

In our case, $m^2 < 0$ and $\gamma > 0$, and the random field has Gaussian white correlation,

$$\langle \eta(t)\eta(t')\rangle_{st} = \varepsilon\delta(t-t'). \tag{20}$$

The equation cannot be solved exactly even within these approximations. If we can neglect the non-linear term $g = 0$, the solution for the Fokker-Plank equation associated with this Langevin equation is given by[12]

$$P(\varphi,t) = (2\pi\varepsilon(t))^{-1/2}\int_{-\infty}^{\infty} dy \exp\left[-\frac{(y - e^{-\gamma t}\varphi)^2}{2\varepsilon(t)}\right]P(y,0) \tag{21}$$

where

$$\varepsilon(t) = \frac{\varepsilon}{\gamma}\left(e^{2\gamma t} - 1\right) \tag{22}$$

and $P(y,0)$ is the initial distribution function at $t = 0$. This form is sufficient for our purpose since the first-stage of SSB is the most relevant to determine the statistical destination of the order parameter $\varphi$. Further in the full non-linear case, only the scaling solution is known[12].

## 3 Dynamics of a system in the environment - evolution from mixed to pure state (process IV)

We now turn our attention to the quantum evolution of the system in the environment. We are interested in how the quantum decoherence and the pro-coherence work. In this section, we study a spin of magnitude half in the environment, which is a part of the complete measurement model we introduce in the next section.

The spin system in the static magnetic field $\vec{B} = (0,0,B)$ and the environment is described by the Hamiltonian.

$$\hat{H} = \hat{H}_S + \hat{H}_I + \hat{H}_B \tag{23}$$

where

$$\begin{aligned}\hat{H}_S &= \omega_0 \hat{S}_3, \\ \hat{H}_I &= g\hat{\vec{S}}\cdot\vec{R},\end{aligned} \tag{24}$$

where $\omega_0 = \mu B$. Since we are interested only in the dynamics of the spin, we coarse grain the environment degrees of freedom. The effective equation of motion is given by[13]

$$\dot{\rho}(t) = -i\omega_0[\hat{S}_3, \rho(t)] + \left(a[\hat{S}_+\rho(t),\hat{S}_-] + b[\hat{S}_-\rho(t),\hat{S}_+] + c[\hat{S}_3\rho(t),\hat{S}_3] + h.c.\right) \tag{25}$$

where the coefficients $a, b, c$ are the reserver correlations

$$\begin{aligned}a &= \frac{g^2}{4}\int_0^\infty dt e^{-i\omega_0 t}\langle \hat{R}_+(t)\hat{R}_-(0)\rangle, \\ b &= \frac{g^2}{4}\int_0^\infty dt e^{i\omega_0 t}\langle \hat{R}_-(t)\hat{R}_+(0)\rangle, \\ c &= g^2\int_0^\infty dt \langle \hat{R}_3(t)\hat{R}_3(0)\rangle,\end{aligned} \tag{26}$$

and are set to be real and positive, discarding imaginary parts which yield irrelevant phase factors[iii]. Note that Eq.(25) is convolution-less local differential equation even after the coarse graining, according to the beautiful formalism [13], which use the pullback of the time evolution. The stationary environment guarantees the fluctuation-dissipation relation

---

include the information of the finite domain size formed after SSB.
[iii] These parameters generally include complex dynamics of the reserver. This approximation is valid only when the evolution of $\rho$ is not too fast and the reserver is stationary.

$$a = \exp[-\hbar\omega_0/(kT)]b \tag{27}$$

where $T$ is the reserver temperature. The above linear equation Eq.(25) is readily solved as

$$\rho(t) = \begin{pmatrix} \dfrac{b(c_1+c_4)+e^{-2(a+b)t}(ac_1-bc_4)}{a+b} & c_2 e^{-(a+b+c+i\omega)t} \\ c_3 e^{-(a+b+c-i\omega)t} & \dfrac{a(c_1+c_4)+e^{-2(a+b)t}(-ac_1+bc_4)}{a+b} \end{pmatrix} \tag{28}$$

for the initial density matrix

$$\rho(t=0) = \begin{pmatrix} c_1 & c_2 \\ c_3 & c_4 \end{pmatrix}. \tag{29}$$

We have taken the matrix representation so that $S_3$ is diagonalized.

Let us consider the high magnetic field (or low temperature) limit $\omega_0 \hbar \gg kT$. It is apparent from the above solution Eq.(28) that the density matrix approaches

$$\rho(t \to \infty) = \begin{pmatrix} \dfrac{b}{a+b} & 0 \\ 0 & \dfrac{a}{a+b} \end{pmatrix}, \tag{30}$$

which represents either the spin-up pure state $|+\rangle = (1,0)$ or the spin-down pure state $|-\rangle = (0,1)$, and not both, depending on the signature of $\omega_0$ (or the direction of the magnetic field) as Eq.(27). Thus the spin state after the contact with the cold environment can become pure, contrary to the ordinary argument emphasizing the decoherence effect of the environment. This fact is demonstrated in Figure 2. The present pro-coherence should be one of the key mechanisms which realize the von Neumann projection in physical systems.

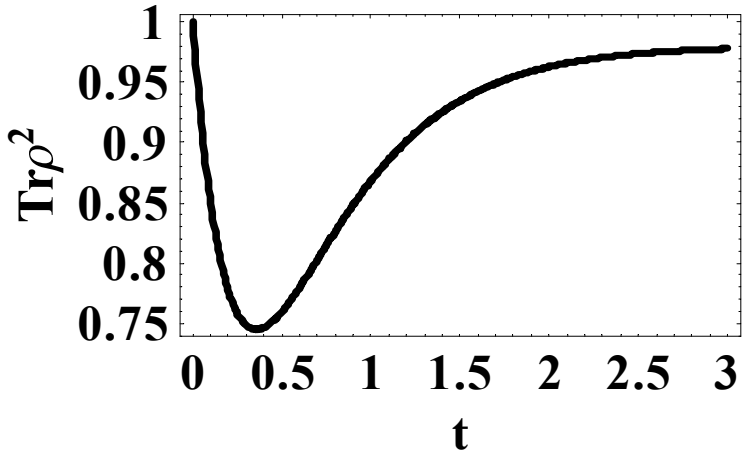

**Figure 2. The solid line represents the time evolution of** $\mathrm{Tr}\rho^2$**, the measure of mixing of the spin state. The state starts from a pure state** $\mathrm{Tr}\rho^2 = 1$**. The system quantum de-coheres (QD) first and then quantum pro-coheres (QP) later. The quantum system in general can become pure from a mixed state after the contact with a cold environment. The parameters in this demonstration are** $c_1$=0.5,$c_2$=0.5,$c_3$=0.5,$c_4$=0.5,a=0.99,b=0.01,c=1,$\omega_0$=10**.**

The above pro-coherence is not restricted to the spin system. In the familiar Caldeira-Legget model [10] for many coupled harmonic oscillators,

$$\hat{H} = \hat{H}_A + \hat{H}_I + \hat{H}_B, \tag{31}$$

$$\hat{H}_A = \frac{\hat{p}^2}{2m} + v(\hat{x}), \tag{32}$$

$$\hat{H}_I = \hat{x}\sum_k C_k \hat{x}_k,$$

the evolution equation for the reduced density matrix can be brought into the convolution-less formalism similar to Eq.(25)[13]. The asymptotic expression of $\text{Tr}\rho^2$ becomes[14],

$$\text{Tr}\rho^2 \xrightarrow[t\to\infty]{} [\text{Coth}(\frac{\hbar\omega}{2kT})]^{-1} \approx \begin{cases} 1 & \text{for } \frac{\hbar\omega}{2kT} \gg 1 \\ \frac{\hbar\omega}{2kT} & \text{for } \frac{\hbar\omega}{2kT} \ll 1 \end{cases} \quad (33)$$

Therefore in the low temperature limit or in the high frequency limit, we have $\text{Tr}\rho^2 \to 1$. Therefore the system becomes a pure state[iv]. Thus in general, a cold environment can make the system pure.

## 4  A simple model of spin measurement

We now introduce the simplest model which satisfies the basic properties of quantum measurement discussed in the introduction. The model is simply a combination of the previously introduce systems in sections 2 and 3, and is composed from a spin of magnitude half (system) and the scalar field with $\lambda\varphi^4$ self interaction (apparatus), with thermal bath. The Lagrangean is given by

$$\hat{L} = \frac{1}{2}(\partial_\mu\hat{\phi})^2 - \frac{m^2}{2}\hat{\phi}^2 - \frac{\lambda}{4!}\hat{\phi}^4 + \mu\hat{\phi}\hat{\vec{S}}\cdot\vec{B} + \text{(bath)} \quad (34)$$

where $\mu > 0$ and the static magnetic field $\vec{B}$ is parallel to the $z$-direction. The wrong sign of the mass term $m^2 < 0$ guarantees the occurrence of SSB, and the c-number order parameter $\varphi$ is described by the effective action $\Gamma[\varphi]$ in the CTP formalism. Although the variable filed $\hat{\phi}$ would not have direct connection to any realistic measuring apparatus, this simple form of Lagrangian will nevertheless be useful to elucidate the relevance of the SSB phase transition for the quantum measurement process.

The evolution of the c-number order parameter can be derived by the variation of the generalized effective action in the CTP formalism. This reduces to the following form similar to Eq.(19) after the slow-rolling and local approximations,

$$\dot{\varphi} = \gamma\varphi - g\varphi^3 + \mu\langle\hat{\vec{S}}\rangle\cdot\vec{B} + \eta . \quad (35)$$

Note that there appears, in the above, an extra term $\mu\langle\hat{\vec{S}}\rangle\cdot\vec{B}$ which biases the SSB phase transition.

On the other hand, the evolution of the system density matrix is given by Eq.(25), but now $\omega_0$ is replaced by $\omega = \mu\varphi(t)B$ and is time dependent through $\varphi(t)$:

$$\dot{\rho}(t) = -i\omega[\hat{S}_3, \rho(t)] + \left(a[\hat{S}_+\rho(t), \hat{S}_-] + b[\hat{S}_-\rho(t), \hat{S}_+] + c[\hat{S}_3\rho(t), \hat{S}_3] + h.c.\right) \quad (36)$$

In the slow rolling regime, the standard fluctuation-dissipation relation Eq.(27) will still hold, and we assume this case. In the limit $\hbar\omega/(kT) \gg 1$, since the magnetic field $B$ is fixed to be constant, the asymptotic state becomes pure spin up state or down state depending now on the signature of the order parameter $\varphi$.

---

[iv] This is most apparent in the representation that $\rho$ becomes diagonal.

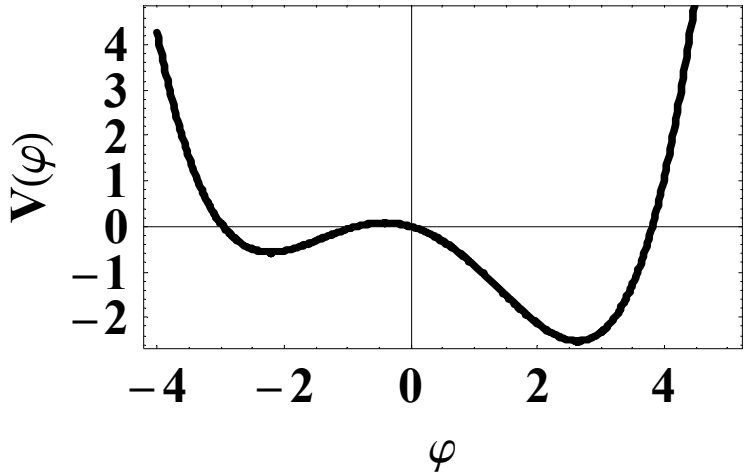

**Figure 3. A biased potential $V(\varphi)$ by the spin coupling $-\mu\varphi\langle\hat{\vec{S}}\rangle\cdot\vec{B}$. In the above case $\langle\hat{\vec{S}}\rangle\cdot\vec{B}>0$ and the slope of the potential at the origin becomes negative. This biases the c-number order parameter field $\varphi$ to roll down toward right. If $\langle\hat{\vec{S}}\rangle\cdot\vec{B}<0$, the potential biases $\varphi$ to roll down toward left.**

Suppose initially the spin is prepared in the state so that $\langle\hat{\vec{S}}\rangle\cdot\vec{B}>0$. Then the *biased* potential $V(\varphi)$ which includes the term $-\mu\varphi\langle\hat{\vec{S}}\rangle\cdot\vec{B}$ has negative gradient at the origin. This promotes the c-number order parameter $\varphi$ to evolve from there toward $\varphi_+>0$. In this evolution, the potential energy for the spin $-\mu\varphi\hat{\vec{S}}\cdot\vec{B}$ favors $\hat{\vec{S}}$ parallel to $\vec{B}$. This is the *positive feed back* for the spin to be *locked* into the up-state with a firm correlation with the order parameter being $\varphi_+$. The biasing toward the other direction and the positive feedback are also true for the initial state so that $\langle\hat{\vec{S}}\rangle\cdot\vec{B}<0$; the spin is locked into the down-state with a firm correlation with the order parameter being $\varphi_-$. This is the essence of the quantum measurement and the von Neumann projection postulate. Due to the strong magnetic field ($\hbar\omega/(kT)\gg 1$), the spin evolves into purely up-state or purely down state in correlation with the c-number order parameter $\varphi_+$ or $\varphi_-$, respectively.

The numerical solutions for the coupled equations Eqs.(35)(36) are depicted in Figure 4.

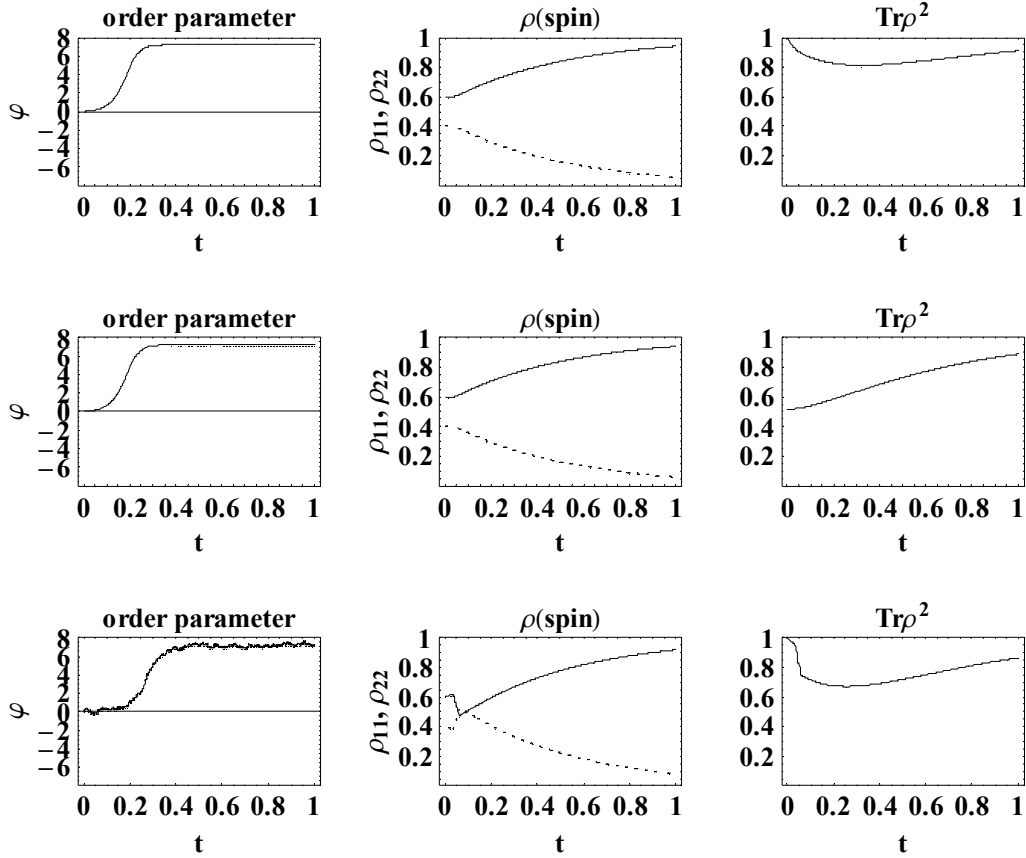

**Figure 4.** Time evolution of the order parameter, spin probabilities($\rho_{11}$ for the solid curve, and $\rho_{22}$ for the broken curve), and the measure of mixing. Fourth order Runge-Kutta method is used with fixed interval, by which the coupling strength of the random field is properly normalized. The first and third rows start from pure states, while the second row starts from a mixed state. Optimization parameters, defined in Eq.(40), are $z = 10, 2.0, 0.08$ respectively for the top to the bottom rows.

The order parameter eventually rolls down the potential toward either $\varphi_+$ or $\varphi_-$ positions and simultaneously the spin state is locked either the up-state or down-state, with firm correlation with the order parameter. In this dynamics, the entropy of the spin system increases first reflecting the QD process and then reduces reflecting the QP process. The initial spin state biases the direction of the rolling order parameter, though the bias is not always faithful because the random force is applied in the SSB process. Though this random field seems to have no relevant role in the present one-event measurement, we will show that the random field plays a crucial role in the realistic ensemble measurement. This is the topic in the next section.

## 5 Efficiency and precision of the measurement

After confirming the above model successfully works as a quantum measurement device, we would like to study to what extent the device actually perform the faithful measurement. For this purpose, the balance between the initial biasing strength and the random force strength turns out to be essential.

The probability that the order parameter is located in the positive $\varphi$ side at time $t$ is given by integrating the solution Eq.(21),

$$P_+ = \int_0^\infty P(\varphi, t) d\varphi = \frac{1 + \mathrm{erf}\left(\delta / \sqrt{2\varepsilon(t)}\right)}{2}, \tag{37}$$

where $\delta = (\mu/\gamma)\langle \hat{\vec{S}} \rangle \cdot \vec{B}$ is the initial bias. The probability that the order parameter is located in the negative $\varphi$ side at time $t$ is likewise given by

$$P_- = \int_{-\infty}^{0} P(\varphi,t)d\varphi = \frac{\text{erfc}(\delta/\sqrt{2\varepsilon(t)})}{2}. \tag{38}$$

In these expressions, the limit $t \to \infty$ would be useless since the solution Eq.(21) is a linear approximation. We have to evaluate these quantities just after the completion of the SSB phase transition. The important time scale for this purpose would be the onset time of order

$$t_0 = \frac{1}{2\gamma}\ln\left[\frac{g}{\gamma}\left(\frac{\varepsilon}{\gamma} + \delta^2\right)\right]^{-1}, \tag{39}$$

which plays an important role in the scaling solution[12]. This expression is the same as the deterministic rolling time duration of the order parameter from the initial position toward the inflection point of the potential; the initial position squared is defined to be the sum of the deviation squared $\delta^2$ and the dispersion $\varepsilon/\gamma$ caused by the random force during the characteristic time scale $1/\gamma$. For the faithful measurement, $P_\pm \approx 1$ should be necessary at $t = t_0$ for the initial pure state $\delta = \mu B/(2\gamma) \equiv \delta_{\max}$. This yields the approximate optimization condition for the measurement[v]:

$$z \equiv \frac{\delta_{\max}}{\sqrt{2\varepsilon(t_0)}} \approx 2. \tag{40}$$

If this optimization condition is satisfied, then the initial expectation value of the system spin along the magnetic field $\langle \hat{\vec{S}}(t=0) \rangle \cdot \vec{B}$ has approximately one-to-one onto correspondence to $P_+(t_0)$. The probability $P_+(t_0)$, in this case, is interpreted to be the relative frequency actually obtained after a large number (ideally, infinite) of measurement on the identically-prepared systems. However this correspondence is not linear in general, and therefore an appropriate calibration would be necessary before the precise measurement. This situation is depicted in Figure 5. If $z \gg 2$, then initial deviation effect dominates the random force and the correspondence becomes almost degenerate in the relative frequency. In this case, an extremely accurate calibration is necessary. On the other hand if $z \ll 2$, then random field dominates the initial deviation effect and the correspondence is no more an onto map; there appears large rage of relative frequency which is never realized. Even in this case, an appropriate calibration still works.

---

[v] The value is somewhat arbitrary since the error function exponentially approaches to one. If we chose too large values, the calibration curve in Figure 5 becomes almost degenerate. Here we have chosen the value $2$ in which case $P_\pm \approx 0.9953$. The following arguments are within this approximation.

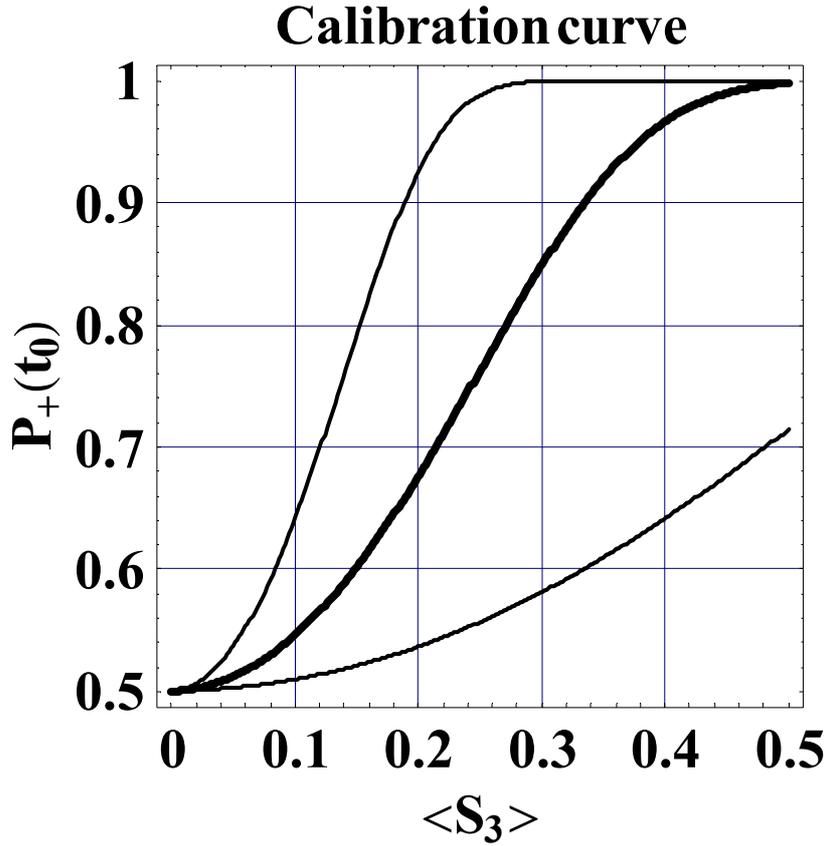

**Calibration curve**

**Figure 5.** The curves represent the probability function evaluated at $t_0$ : $P_+(t_0)$ as a function of the initial expectation value of the system spin $\hat{S}_3$. For about the spin variable $\langle \hat{S}_3 \rangle$, only the half of the full range $[-1/2, 1/2]$ is shown above because of the curves have rotational symmetry of degree $\pi$ around $(0, 0.5)$. The probability $P_+(t_0)$ is considered to be a relative frequency actually obtained after a large number of measurement on the identically-prepared systems. The thick curve represents the most optimized case with $z = 2.0$. The thin curve below is for $z \ll 2$. Even in this case, an appropriate calibration makes this measurement apparatus faithful. The thin curve above is for $z \gg 2$. For this case, an extremely accurate calibration would be necessary since the relative frequency becomes almost degenerate.

Though the random field plays a crucial role in the first stage of the SSB phase transitions as we have seen above, it becomes harmful in the later stage of the SSB phase transition. Actually it induces fluctuations of $\varphi$ from the stationary values $\varphi_\pm$, and reduces the sharpness of the pointer readout, though this effect would be small. The random field is preferable only in the first stage of SSB.

## 6 Conclusions and some applications

Our original goal has been to clarify the physical process of the von Neumann projection, and resolve the original double-structure in the evolution rules in quantum mechanics. We first introduced the hypothesis that the crucial process in the quantum measurement is the spontaneous symmetry breaking (SSB). The demonstration of this hypothesis requires us to study the description of the c-number order parameter in the CTP quantum field theory and the pro-coherence in the effective spin dynamics. Combining these basic considerations, we have introduced a measurement model with the spin-field coupling. While the apparatus+spin system establishes the entanglement (QE) with each other during their contact, the random fields acting from the outside environment provide the quantum decoherence (QD) of the spin system as well as the trigger of the phase transition (SSB) of the apparatus. This measurement model with SSB guarantees the positive feedback of the evolution of the order parameter to the spin evolution, which promotes the pro-coherence (PC) of the spin system. *These four processes, QE, QD, SSB, and PC form the basic*

*quartet of the quantum measurement.*

We have further considered the efficiency and the precision of the measurement on an ensemble of identically-prepared systems. We have obtained the *calibration curve*, by which we can predict the initial spin state, and the *optimization condition* for the most efficient measurement. The necessity of this calibration and the characteristic time scale of the SSB phase transition $t_0$ in Eq.(39) are characteristics of our approach and also the check points for the validity of our scenario in comparison with actual quantum measurements.

We have, so far, considered the simplest SSB of discrete $Z_2$ symmetry in the measurement model. Further application is of course possible to the SSB models with continuous symmetry. For example, consider the measurement of the photon phase $\hat{\omega}$. We prepare the potential with the following coupling between the photon field $a$ and the apparatus $\phi$ which is now a complex scalar filed,

$$V(\hat{\phi}) = -\mu^2 \hat{\phi}^\dagger \hat{\phi} + \frac{\lambda}{2}(\hat{\phi}^\dagger \hat{\phi})^2 + \mu(\hat{\phi}^\dagger \hat{a} + \hat{a}^\dagger \hat{\phi}) + O(\hbar). \tag{41}$$

The original $U(1)$ symmetry is broken after the evolution of the c-number order parameter from 0 toward finite values $\varphi = |\varphi|e^{i\theta}$ ($|\varphi| : 0 \to \sqrt{2\mu^2/\lambda}$) with a spontaneously selected phase $\theta$. During this process the photon phase is strongly locked into $\omega \approx \pi - \theta$ with positive feedback. Thus this system works as a proper quantum measurement apparatus. Also here the basic quartet is the essence of the quantum measurement. However the appearance of the Goldstone mode may enhance the fluctuation effect by the random force, and therefore not preferable for the exact measurement.

In the above models, we have studied the projective (ideal) measurements. Similar modeling is possible also for the absorptive measurements, in which the system is absorbed and the quantum information of the system is lost while the c-number order parameter newly appears. Suppose the field $\phi(t, \vec{x})$, characteristic of the apparatus, extends its support in the spatial domain $\mathfrak{D}$ in the symmetric phase. This means that the meta-stable spatial region $\mathfrak{D}$ is ready to accept the injection of a system. A local phase transition reaction in some place in $\mathfrak{D}$ completes the measurement. As a result, a special location in $\mathfrak{D}$ is spontaneously selected which characterize the injection of a system particle. In this case, the process PC is missing among the basic quartet. There are many other applications such as the negative result experiments, quantum Zeno effects, etc. which will be reported in our future publications.

Before concluding our paper, we would like to add two basic considerations.

One is the application of our formalism to the Universe. Originally the resolution of the double structure in quantum mechanics has been strongly desired in the field of cosmology. Especially in the early Universe, we have to clarify the quantum origin of the density fluctuations after the complete level out by the inflationary mechanism[2]. The field $\varphi$ in our model plays a very similar role to the inflaton field. In this cosmological context, we recognize that the relevant quantity is the c-number order parameter, which spontaneously breaks the translational invariance, but not the quantum state after some kind of quantum measurement. In this sense, the problem resembles that in the absorptive measurement in the lack of QP process among the basic quartet.

Another is related with the basic Born's probability rule in quantum mechanics: After the measurement of the observable $\hat{A}$ in the state $\psi \in \mathcal{H}$, the obtained value is one of the eigenvalues of $\hat{A}$ with the probability $P(a)$ given by $P(a) = \|\mathcal{P}_a |\psi\rangle\|^2$. In our argument in section 5, we did obtain some information on the probability as well as the von Neumann projection. If we adopt the most optimized case with the appropriate calibration, the probability is completely predicted by the thick curve in Figure 5. In this sense, our measurement model claims more than a simple projection. Furthermore if the parameters satisfy $\gamma \gg \sqrt{g\varepsilon}$ (strong friction) or $\varepsilon \gg \gamma\delta^2$ (strong random force), then we can have the linear relation for the spin-up probability $P_+ = \frac{1}{2} + \alpha\left(2\|\mathcal{P}_+|\psi\rangle\|^2 - 1\right)$, where $\alpha$ is a real constant. Moreover if $\alpha = 0.5$, then we have indeed obtain the result $P_+ = \|\mathcal{P}_+|\psi\rangle\|^2$, the exact form of the Born's probability rule.

Finally we conclude our paper with mentioning much profound problems. Does the SSB mechanism always associate with the quantum measurement process? How the exact derivation of the dynamics of SSB phase transitions is possible? Our present model of quantum measurement may also be the emerging mechanism of the macroscopic irreversibility. The problem is whether SSB phase transition is always associated with the emergence of macroscopic irreversibility. These problems should be addressed sometime in future for the complete resolution of the issue associated with the quantum measurement.

# Acknowledgement

One of the authors (MM) would like to thank Akira Shimizu, Masahiro Hotta, Fumiaki Shibata and Akio Hosoya for fruitful discussions and valuable comments.